# Controlling the Electrical Properties of Undoped and Ta-doped TiO$_2$ Polycrystalline Films via Ultra-Fast Annealing Treatments

*Piero Mazzolini, Tolga Acartürk, Daniel Chrastina, Ulrich Starke, Carlo Spartaco Casari, Giuliano Gregori*[†] *and Andrea Li Bassi*[†]

P. Mazzolini
1) Department of Energy - Politecnico di Milano, via Ponzio 34/3, I-20133 Milano, Italy. 2) Center for Nano Science and Technology - IIT@PoliMI, Via Pascoli 70/3, I-20133 Milano, Italy.

T. Acartürk
3) Max Planck Institute for Solid State Research, Heisenbergstr. 1, 70569 Stuttgart, Germany.

Dr. D. Chrastina
4) L-NESS, Dipartimento di Fisica, Politecnico di Milano - Polo Territoriale di Como, Via Anzani 42, I-22100 Como, Italy

Prof. Dr. U. Starke
3) Max Planck Institute for Solid State Research, Heisenbergstr. 1, 70569 Stuttgart, Germany.

Prof. Dr. C. S. Casari
1) Department of Energy - Politecnico di Milano, via Ponzio 34/3, I-20133 Milano, Italy. 2) Center for Nano Science and Technology - IIT@PoliMI, Via Pascoli 70/3, I-20133 Milano, Italy.

Dr. G. Gregori
3) Max Planck Institute for Solid State Research, Heisenbergstr. 1, 70569 Stuttgart, Germany.
E-mail: g.gregori@fkf.mpg.de
tel.: +497116891770

Prof. Dr. A. Li Bassi
1) Department of Energy - Politecnico di Milano, via Ponzio 34/3, I-20133 Milano, Italy. 2) Center for Nano Science and Technology - IIT@PoliMI, Via Pascoli 70/3, I-20133 Milano, Italy.
E-mail: andrea.libassi@polimi.it
tel.: +390223996316

[†] The two authors contributed equally to the manuscript

We present a study on the crystallization process of undoped and Ta doped TiO$_2$ amorphous thin films. In particular, the effect of ultra-fast annealing treatments in environments characterized by different oxygen concentrations is investigated via *in-situ* resistance measurements. The accurate examination of the key parameters involved in this process allows us to reduce the time needed to obtain highly conducting and transparent polycrystalline thin films (resistivity ~ $6 \times 10^{-4}$ Ωcm, mean transmittance in the visible range ~ 81%) to just 5 minutes (with respect to the 180 minutes required for a "standard" vacuum annealing treatment) in nitrogen atmosphere (20 ppm oxygen concentration) at ambient pressure. Experimental evidence of superficial oxygen incorporation in the thin films and its detrimental role for the conductivity are obtained by employing different concentrations of traceable $^{18}$O isotopes during ultra-fast annealing treatments. The results are discussed in view of the possible implementation of the ultra-fast annealing process for TiO$_2$-based transparent conducting oxides as well as electron selective layers in solar cell devices; taking advantage of the high control of the ultra-fast crystallization processes which has been achieved, these two functional layers are shown to be obtainable from the crystallization of a single homogeneous thin film.

**1. Introduction**

TiO$_2$ is one of the key materials for energy applications such as lithium-ion batteries, photocatalysis, water splitting, and charge carrier separation/collection in solar cells.[1] If we consider photovoltaic applications, TiO$_2$ in the anatase phase is the most employed material for several architectures of dye sensitized and perovskite-based solar cells:[2-4] this is mainly related to a high chemical stability, its intrinsic transparency to visible light (bandgap $E_g$ = 3.2 - 3.4 eV[5, 6]) and to the favorable energy level alignment at the solar cell interfaces, which enables an efficient and selective charge carrier uptake and transport (n-type charge transport as a photoanode and/or as a hole-blocking, electron selective layer) from the photoactive material to the front transparent electrode of the solar cell.[7, 8] This electrode is generally a thin film of a transparent conducting oxide (TCO), usually fluorine-doped tin oxide (FTO).[9]

However, the discovery of donor doped $TiO_2$ as a promising new class of n-type TCO,[10] and the possibility of obtaining highly conducting and transparent polycrystalline Nb or Ta-doped $TiO_2$ (TaTO) thin films on inexpensive glass substrates, open up new opportunities for a better suited energy level alignment at the device interfaces via an all $TiO_2$-based configuration (photoanode/selective layer and TCO).[11-13] Between the two mentioned donor doped compositions, Ta-doped $TiO_2$ is thought to have several advantages with respect to the more widely investigated Nb for transparent conductor applications, namely higher mobility and dopant solubility.[14] Moreover, it has been suggested that donor-doped $TiO_2$ could be a superior material compared to undoped $TiO_2$ not only as a TCO but also as an electron selective layer (i.e. hole blocking layer) and photoanode.[15-17] In both cases the chosen material must provide efficient electron transport pathways, combined with a low charge carrier density, in order to reduce the recombination rate with the photogenerated charges.[18] It is important to mention that depending on the adopted synthesis conditions (reducing vs. oxidizing deposition/annealing atmosphere) it is possible to finely tune the mobile charge carrier density in $TiO_2$-based films as well as their functional (electrical/optical) properties.[12, 15] Typically, in order to achieve the highest electrical conduction upon deposition, $TiO_2$-based thin films require an annealing process in a reducing atmosphere (vacuum or $H_2$-based atmosphere, commonly at temperatures between 500 °C and 600 °C) so as to induce crystallization to a pure anatase phase and thus increase the carrier mobility, without compromising the charge carrier density.[12, 13, 19] In this context, it is noteworthy that annealing processes in oxidizing atmospheres (e.g. air) result in highly insulating films.[12] Although such a result is expected from defect chemistry considerations, the exact mechanisms underlying this process and involving different phenomena (crystallization, oxygen incorporation) are still debated especially in the case of donor-doped $TiO_2$.[20-27]

In this contribution, we focus on how the post-deposition annealing treatments performed in different atmospheres and different heating/cooling rates affect the electrical properties of undoped as well as tantalum doped anatase TiO$_2$ thin films. More importantly, we investigate the role of the change of the microstructure (from amorphous to crystalline) and of the oxygen exchange with the surroundings on the final electrical conductivity of the material. For this purpose, we consider amorphous films grown via pulsed laser deposition (PLD) at room temperature on soda lime glass substrates. We show that ultra-fast-annealing treatments (UFA: heating rate of 300 K/min up to the peak temperature of 460 °C and total treatment time of ~ 5 minutes) performed in a N$_2$ atmosphere (with 20 ppm O$_2$ concentration) at ambient pressure allow high quality transparent and conducting anatase thin films to be obtained. In particular, UFA-treated tantalum doped (TaTO) films show practically identical electrical and optical properties ($\rho_{min}$ ~ 6 × 10$^{-4}$ Ωcm, transmittance in the visible range T$_{VIS}$ ~ 81%) as TaTO layers, which were treated according to a vacuum annealing cycle which was already shown to be effective for obtaining the best optical and conduction properties (in this work this will be referred to as "standard annealing cycle": $p < 4 \times 10^{-5}$ Pa, overall time heating + dwell at 550 °C + cooling ~ 180 minutes[12]). It is worth mentioning that a previous study on Nb-doped TiO$_2$ thin films reported on the possibility of obtaining good resistivity values ($\rho = 8.4 \times 10^{-4}$ Ωcm) upon annealing in a diluted atmosphere (0.5 atm) of highly pure nitrogen (purity 99.9998%, nominal oxygen concentration < 0.5 ppm) at 350 °C for 20 minutes.[28] The physicochemical mechanisms behind this result (e.g. role of possible oxygen incorporation during annealing) were however not discussed.

Here instead in-situ electrical measurements carried out during UFA enable us to study the thin film crystallization process, and to identify the threshold limit of oxygen concentration for which the electrical properties start to be negatively affected for Ta-doped and undoped TiO$_2$. In order to investigate the possible oxygen incorporation/diffusion into the films during annealing, we perform time-of-flight secondary ion mass spectrometry (TOF-SIMS) on TaTO samples crystallized using UFA under different $^{18}$O concentrations (80 ppm and 1000 ppm). This also allows us to deconvolve the role of the crystallization process from the annealing atmosphere on the final electrical properties of the films. Finally, the findings are discussed also in view of the possible technological implications in the field of next generation solar cells.

## 2. Materials and Methods

### 2.1 Thin Film Deposition

Amorphous Ta-doped $TiO_2$ (TaTO) and $TiO_2$ thin films were grown by room temperature PLD on soda-lime glass ($10 \times 10 \times 1$ mm$^3$) and Si (100) substrates at room temperature in the presence of an oxygen background pressure of about 1 Pa (TaTO was deposited at 1 Pa, while $TiO_2$ at 1.25 - 1.3 Pa). Ablation was performed from $TiO_2$ (powder purity 99.9%) or $Ta_2O_5$:$TiO_2$ (molar ratio of 0.025:0.975, powder purity 99.99%) targets to deposit $TiO_2$ or TaTO films, respectively. A ns-pulsed Nd:YAG laser (4$^{th}$ harmonic, $\lambda = 266$ nm) with a repetition rate $f_p = 10$ Hz, a pulse duration of ~ 6 ns and a laser fluence of 1.15 J/cm$^2$ was used. The target-to-substrate distance was fixed at 50 mm.

### 2.2 Structural Characterization

The film thickness was evaluated by means of scanning electron microscopy (Zeiss SUPRA 40 field-emission SEM) on samples grown on silicon. The crystalline structure was determined by X-ray diffraction (PANalytical X'Pert PRO MRD high-resolution X-ray diffractometer, using CuK$_{\alpha1}$ radiation ($\lambda = 0.15406$ nm) selected by a two-bounce Ge monochromator). XRD measurements were performed in both $\theta$-$2\theta$ and grazing incidence angle configuration (fixed incidence angle $\omega = 5°$). The thin film surfaces were investigated by means of optical microscopy using polarized light (Leica DM2500 M). Optical transmittance spectra (in the range 250 - 2000 nm) were evaluated with a UV–vis–NIR PerkinElmer Lambda 1050 spectrophotometer with a 150 mm diameter integrating sphere. All the acquired spectra were normalized with respect to the glass substrate contribution.

**2.3 Thin Film Crystallization**

As-deposited $TiO_2$ and TaTO thin films were annealed in a standard vacuum ($p < 4\times10^{-5}$ Pa, obtained with an Agilent Varian Turbo-V 250 Turbomolecular Pump) or nitrogen (99.999 % purity, oxygen concentration nominally < 3 ppm) atmosphere (1 atm, obtained after previous vacuum at $p < 4\times10^{-5}$ Pa) in a home-made furnace at 550°C (10 K/min ramps) for 1 hour and used as reference samples, according to reference[12]. The resistivity, mobility and charge carrier density were evaluated via Hall measurements (DC 4-point probe configuration) at room temperature with a Keithley K2400 Source/Measure Unit as a current generator (from 100 nA to 10 mA), an Agilent 34970A voltage meter, and a 0.57 T Ecopia permanent magnet. The crystallization process of the amorphous $TiO_2$-based thin films was studied with ultra-fast annealing (UFA) consisting of ultra-fast temperature ramps (300 K/min) up to the peak temperature of 460 °C (without dwell time) in a home-made furnace employing 5 IR lamps (RS Components Ltd. UK, Heat lamp 500 W R7s 230 V) chosen to avoid any UV emission, concentrically placed outside the chamber. The UFA treatments were performed at ambient pressure under different oxygen/nitrogen mixtures (oxygen concentrations: 20 ppm, 1000 ppm and 21%, which were monitored with a Cambridge Sensotec RapidOX 2100ZF lambda sensor). The 20 ppm gas was obtained by directly employing nitrogen from the lab distribution line, while the 1000 ppm and 21% mixtures were obtained by properly mixing $N_2$ and $O_2$ from a 5.0 purity oxygen bottle. The background gas was continuously flowing in the chamber with a fixed flux of 50 sccm. The temperature was measured using a K-type thermocouple (diameter 0.5 mm) placed on the sample holder at 0.5 mm from the substrate. The furnace design allows also for a fast cooling ramp (about 150 K/min), which can be achieved by an external flux of cold $N_2$ gas. The total treatment time (heating + cooling) required for a typical UFA experiment was around 5 minutes. Changes of the thin film electrical resistance were recorded *in-situ* through 2-point DC measurements (source/measure unit Keithley 2604B). The maximum applied current was 1 mA and the maximum compliance voltage 5 V. In order to minimize the effect of different contact geometries among different samples, Ti/Au (200/2000 Å) electrodes with 1 mm distance from each other were evaporated on top of the amorphous films. All the analyzed samples showed an ohmic behavior when electrically measured between the two evaporated contacts. UFA treatments were also performed on both TaTO and $TiO_2$ thin films without employing evaporated electrodes, in order to subsequently measure their optical and electrical properties via *ex-situ* 4-point configuration Hall measurements with contacts placed on the top surface. This allowed for a direct comparison of these samples with those which are prepared according to the standard annealing process.

**2.4 Tracing of Oxygen Incorporation**

In order to trace the oxygen incorporation we performed UFA on TaTO thin films in $^{18}O$ containing $N_2$ atmospheres (~80 ppm and ~1000 ppm, respectively). The $^{18}O/^{16}O$ concentration profiles as a function of depth were obtained by Secondary Ion Mass Spectrometry, using a commercial TOF-SIMS IV. The secondary ions were generated by short pulses of a 25 keV Ga ion beam. Removal of material for depth profiling was carried out using a second ion beam from a Cs source operated at 500 eV. The correlation between the secondary ion extraction and depth was established using a Dektak 8 profilometer at the end of the measurement.

The possible presence of an insulating top layer corresponding to superficial oxygen penetration was investigated by removing the first tens of nm of an UFA-treated TaTO film by bombardment by Ar ions at 0.2 kV accelerating voltage. The sputtering time was calibrated by SEM images.

**3. Results**

Room temperature PLD of both TiO$_2$ and TaTO yields thin films characterized by absence of long range crystalline order, as discussed in our previous work.[12] The thickness of all the analyzed films is 150 +/- 5 nm, as measured by SEM. The resistivity of all the as-deposited thin films is about 10 Ωcm.

**3.1 Standard Annealing Process**
A standard post deposition annealing process (i.e. $T$ = 550 °C for 1 hour with heating and cooling rates of 10 K/min) was performed on undoped and Ta-doped TiO$_2$ amorphous films in vacuum, since this is a well-established procedure for obtaining highly conducting and transparent polycrystalline anatase films.[12] The room temperature resistivity of TaTO thin films was one order of magnitude lower with respect to TiO$_2$ ($\rho$ = 6.77 × 10$^{-4}$ Ωcm and 6.93 × 10$^{-3}$ Ωcm respectively). This is not surprising, as Ta substituting Ti in the anatase cell is an electron donor, resulting in a larger concentration of mobile electrons ($n$ = 7.99 × 10$^{20}$ cm$^{-3}$ and 5.70 × 10$^{19}$ cm$^{-3}$ for TaTO and TiO$_2$ respectively).[20] On the other hand, the room temperature mobility value is slightly higher for the undoped sample ($\mu$ = 11.5 cm$^2$V$^{-1}$s$^{-1}$ and 15.8 cm$^2$V$^{-1}$s$^{-1}$ for TaTO and TiO$_2$ respectively). Nonetheless, one should note here that both values belong to the best mobility range for TiO$_2$-based polycrystalline films.[12] The mean transmittance values in the visible range ($T_{VIS}$ evaluated in the range 400 – 700 nm) are both within the important technological range for transparent electronics of 80% (81.4% and 79.3% for TaTO and TiO$_2$ respectively).[29]

The TiO$_2$-based thin films were also annealed using a standard annealing process in 1 atm of pure nitrogen gas (grade 99.999 %, nominal O$_2$ concentration < 3 ppm). The resulting resistance of the thin films treated in N$_2$ was however too large for our experimental setup (maximum applicable current of 0.1 μA, voltage compliance 10 V) and thus no value of the resistivity could be determined.

**3.2 Ultra-Fast Annealing**

Since an abrupt drop in the electrical resistivity during the heat treatment of amorphous $TiO_2$-based thin films was already proposed to be a sign of its crystallization,[13, 27] the analysis of the resistance behavior of the thin films recorded via *in-situ* electrical measurements during the UFA treatments permits the threshold temperature and the time needed to crystallize the amorphous thin films in the presence of different oxygen concentrations to be identified. The acquired data for doped and undoped $TiO_2$ samples are plotted in **Figure 1** and **Figure 2** (electrical resistance represented by dots linked to left y-axes, measured temperature represented by dashed lines linked to right y-axes both as a function of time).

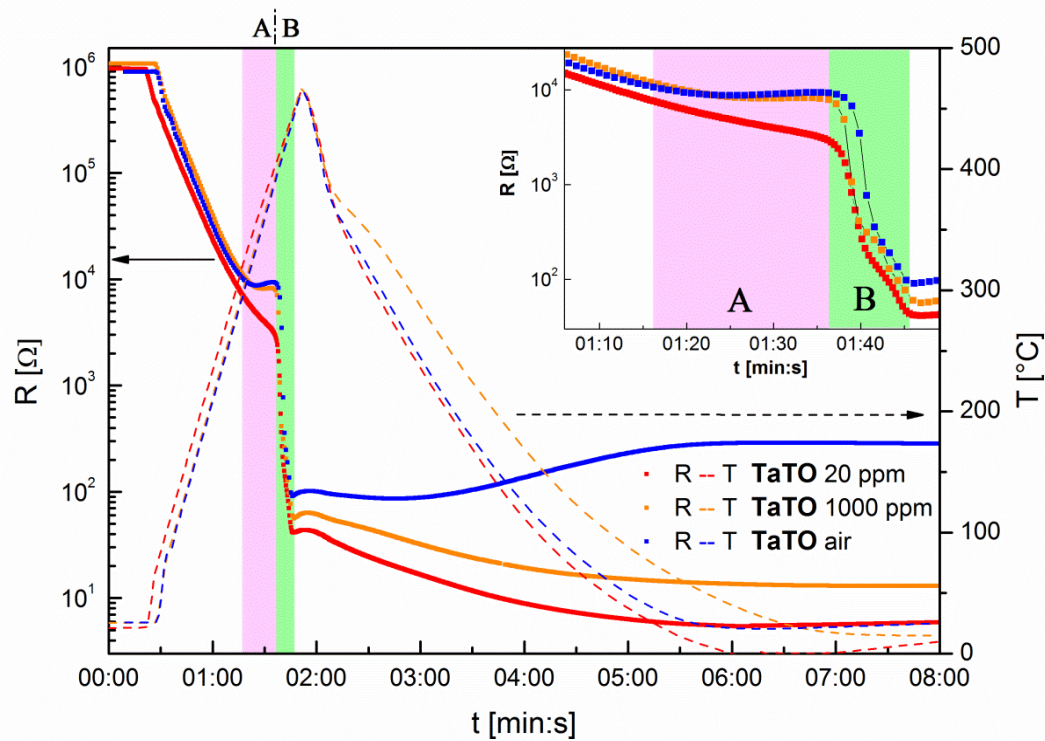

**Figure 1.** In-situ resistance measurements (dots, left y-axes) and corresponding temperature profiles (dashed lines, right y-axes) for TaTO thin films crystallized in $N_2$-based atmospheres with different oxygen concentrations: 20 ppm (red), 1000 ppm (orange) and 21% (synthetic air, blue) atmospheres. The regions A and B (pink and green colored regions respectively) represent the time intervals in which the resistivity of the thin films starts to be affected by the presence of different $p_{O2}$ (A) and the time intervals in which the abrupt resistance drop takes place (B). In the inset is reported a magnification of the resistance behavior in the regions A and B.

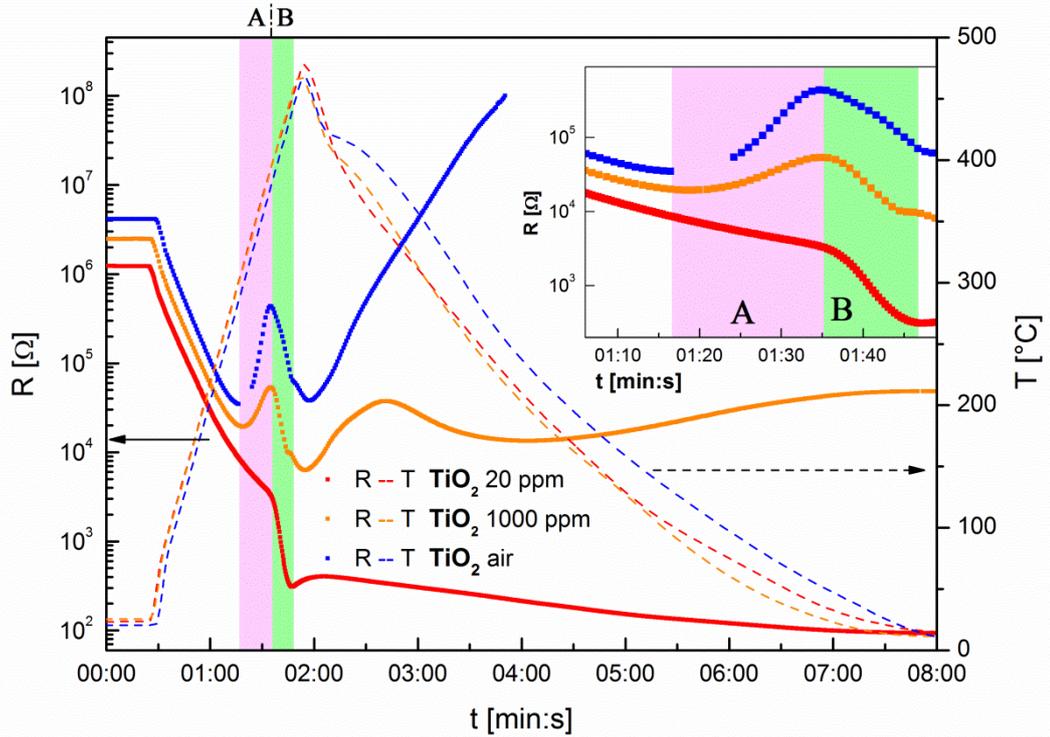

**Figure 2.** In-situ resistance measurements (dots, left y-axes) and corresponding temperature profiles (dashed lines, right y-axes) for $TiO_2$ thin films crystallized in $N_2$-based atmospheres with different oxygen concentrations: 20 ppm (red), 1000 ppm (orange) and 21% (synthetic air, blue) atmospheres. The regions A and B (pink and green colored regions respectively) represent the time intervals in which the resistivity of the thin films starts to be affected by the presence of different $p_{O2}$ (A) and the time intervals, in which the abrupt resistance drop takes place (B). In the inset is reported a magnification of the of the resistance behavior in the regions A and B. There are no resistance points for an electrical resistance higher than $1 \times 10^8$ Ω (blue dots) since this is the upper limit measurable with our experimental setup.

Irrespective of the presence of the dopant or the exposure to different annealing atmospheres, all the performed UFA treatments show an abrupt and sharp decrease of the resistance once a temperature of around 400 °C is reached. As was already mentioned, this transition is likely to be connected to the crystallization process of the amorphous films, the occurrence of which is highlighted in Figure 1 and 2 by the area in green (called region-B).[13] Interestingly, the temperature and time values defining this green zone ($T_{start}$, $T_{end}$ and $t_{start}$, $t_{end}$) appear to be very similar for $TiO_2$ and TaTO thin films: the $T_{start}$ and $T_{end}$ of this process are at around 400 °C and 450 °C respectively ($t_{start}$ - $t_{end}$ = 10 s) for all the investigated samples. It is possible to estimate the abrupt drop in the electrical resistance in region-B to be on the order of a factor of 10 for TiO$_2$, and 100 for TaTO, although here the temperature dependence of the electrical resistance (d$R$/d$T$) is not taken into account ($\Delta T$ ~ 50 °C).

Remarkably, the UFA performed in N$_2$ atmosphere (with 20 ppm O$_2$, red curves in Figure 1 and 2), resulted in low resistance TiO$_2$-based thin films showing a metallic behavior in the cooling region (d$R$/d$T$ > 0). The obtained room temperature resistance ($R_{T=25°C}$) for TaTO is one order of magnitude lower than for TiO$_2$ (6 Ω and 96 Ω, respectively). It is important to note that this difference is of the same order of magnitude of that obtained upon standard annealing treatments carried out in vacuum.

Nevertheless, a substantial difference in the resistance behavior during the heating cycle, also confirmed by the resistance values measured at room temperature, is observed when the UFA experiment is carried out in a more oxygen rich environment. In the case of TaTO, UFA performed in an N$_2$-based atmosphere with 1000 ppm O$_2$ (orange curves in Figure 1), resulted in a more than doubled $R_{T=25°C}$ (13 Ω with respect to the 6 Ω obtained in 20 ppm of oxygen). Nevertheless the thin film maintained a metallic behavior during cooling as shown in Figure 1 for $t$ > 2 min. In contrast, the crystallization process performed in the same conditions for TiO$_2$ resulted in a semiconducting behavior during cooling (d$R$/d$T$ < 0, orange curves in Figure 2 for $t$ > 4 min) while the R$_{T=25°C}$ is orders of magnitude higher than for the UFA treatment carried out under 20 ppm of oxygen (4.2 × 10$^4$ Ω vs. 96 Ω). The effect of a further increase of the oxygen concentration is shown for UFA performed in an artificial air atmosphere (21% O$_2$ in N$_2$, blue curves in Figure 1 and 2). The electrical properties of both TaTO and TiO$_2$ thin films are significantly different compared to those of the films which were ultra-fast-annealed in 20 and 1000 ppm of oxygen. As a matter of fact, the resulting $R_{T=25°C}$ was increased up to 278 Ω for TaTO, while for TiO$_2$ it was too high to be measurable with our experimental setup ($R$ > 10$^8$ Ω).

We found that the increasing loss of conductivity with increasing O$_2$ concentration was associated with a different resistance behavior recorded before the abrupt resistance drop in region-B. This is evident in the A-regions (highlighted in pink) in the UFA graphs shown in Figure 1 and 2. In fact the resistance change with increasing temperature for the amorphous samples treated under 20 ppm $O_2$ (red curves in Figure 1 and 2) could be defined as an almost-monotonic decrease of resistance until the abrupt drop begins (region-B). In contrast, for UFA performed under higher oxygen partial pressures (1000 ppm and air – 21%, orange and blue curves respectively in Figure 1 and 2), a resistance increase in the region-A is recorded. Moreover, it is interesting to note that the time interval (and consequently also the $\Delta T$) associated with this phenomenon is again similar for TaTO and $TiO_2$ thin films, although this effect is definitely much more pronounced for the undoped samples ($t_{start}$ - $t_{end}$ ~ 20 s; $T_{start}$ ~ 300 °C, $T_{end}$ ~ 400 °C).

Extended UFA treatments were performed in order to investigate the effect on the resistance of a longer exposure (10 min) to 20 ppm $O_2$ at the peak temperature (460 °C). Note that all the other parameters were kept constant (ambient pressure, 20 ppm $O_2$ in $N_2$, and the same heating and cooling rates as in the conventional UFA treatments). The results, which are shown in the Supporting Information (**Figure s1**), clearly indicate that a longer exposure to $O_2$ at 460 °C led to the degradation of the electrical conductivity of both TaTO and $TiO_2$ thin films ($R_{T=25°C}$ = 11 Ω and 4965 Ω for TaTO and $TiO_2$ respectively) with respect to the conventional UFA cycle ($R_{T=25°C}$ = 6 Ω and 96 Ω for TaTO and $TiO_2$ respectively).

We also performed the UFA treatments on samples without electrodes, in order to properly compare their optical and electrical properties (4-point van der Pauw – Hall resistivity measurements) with the samples prepared by the standard annealing process performed under vacuum and $N_2$. Once annealed, the room temperature electrical properties obtained for TaTO and $TiO_2$ thin films are stable in time (samples measured over several months), regardless of the different temperature cycle employed (standard or UFA). UFA treatment of TaTO in 1 atm of hydrogen-containing atmosphere (Ar/$H_2$ mixture, $H_2$ at 2% - measured oxygen concentration < $10^{-20}$ ppm) was performed and compared to the standard vacuum annealing process.

**Table 1.** Electrical properties of TaTO and TiO$_2$ thin films evaluated via 4-point electrical measurements at room temperature. The "/" symbol means that it was not possible to obtain a reliable experimental value due to the highly scattered collected data. The "n. m." abbreviation indicates that the samples were too insulating to be measurable with our experimental setup. The background colors used in the rows referring to N$_2$-based UFA treatments correspond to those in Figure 1 and 2.

|  | **Annealing treatment** | **$\rho$ [$\Omega$cm]** | **$n$ [cm$^{-3}$]** | **$\mu$ [cm$^2$V$^{-1}$s$^{-1}$]** |
|---|---|---|---|---|
| **TaTO** | **Standard-Vacuum** | $6.77 \times 10^{-4}$ | $7.99 \times 10^{20}$ | 11.5 |
|  | **Standard-N$_2$** | n. m. | n. m. | n. m. |
|  | **UFA – Ar/H$_2$** | $7.65 \times 10^{-4}$ | $7.46 \times 10^{20}$ | 11.0 |
|  | **UFA – 20 ppm** | $6.85 \times 10^{-4}$ | $8.30 \times 10^{20}$ | 11.0 |
|  | **UFA – 1000 ppm** | / | / | / |
|  | **UFA - air** | / | / | / |
| **TiO$_2$** | **Standard Vacuum** | $6.93 \times 10^{-3}$ | $5.70 \times 10^{19}$ | 15.8 |
|  | **Standard N$_2$** | n. m. | n. m. | n. m. |
|  | **UFA – 20 ppm** | $9.04 \times 10^{-3}$ | $4.77 \times 10^{19}$ | 14.5 |
|  | **UFA – 1000 ppm** | n. m. | n. m. | n. m. |

Notably, the ultra-fast-annealed TaTO film in reducing atmosphere (Ar/H$_2$) and the film crystallized in 20 ppm O$_2$ have electrical properties almost identical to those obtained with a standard vacuum annealing (see **Table 1**). In the case of TiO$_2$, the UFA treatment performed in 20 ppm of oxygen resulted in a slightly higher resistivity, although the overall electrical properties are comparable. On the other hand, a standard annealing process performed in a highly pure N$_2$ atmosphere (O$_2$ < 3 ppm) results in an insulating film ($\rho$ not measurable). Moreover, it is worth mentioning that in the case of UFA treatments performed with O$_2$ concentrations of 1000 ppm and 21% on TaTO samples without electrodes, a notable increase of the resistivity with respect to the best recorded values was recorded ($3.39 \times 10^{-3}$ $\Omega$cm and $2.31 \times 10^{-2}$ $\Omega$cm for 1000 ppm and 21% respectively). Nonetheless, these data have to be considered as only indicative due to the diode-like *IV* characteristics between the electrical probes and the sample, which invalidates the 4-point measurement method, and are consequently not reported in Table 1. On the other hand, in the case of $TiO_2$, a UFA treatment in presence of 1000 ppm of oxygen was sufficient to result in an insulating film ($\rho$ not measurable).

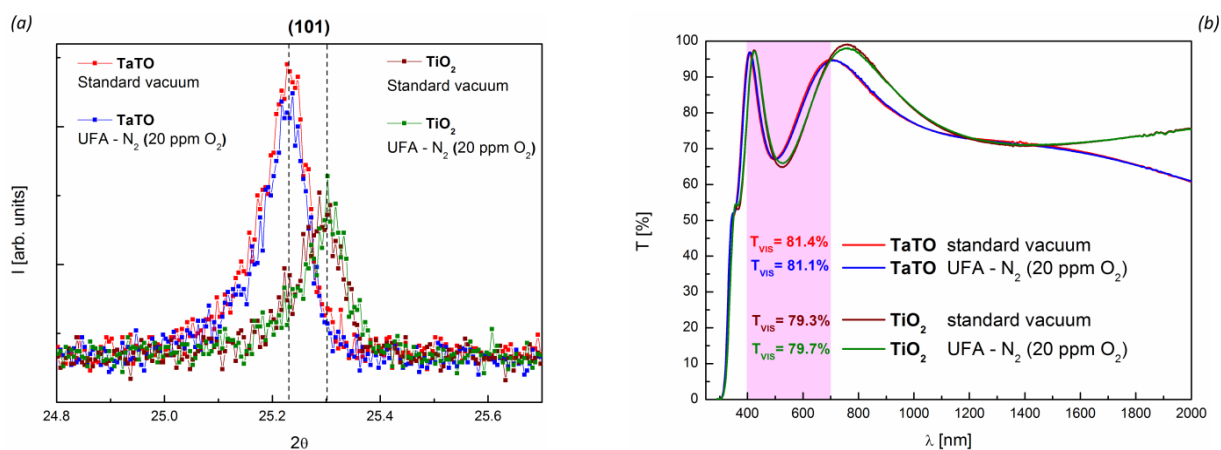

**Figure 3.** *(a)* the XRD acquisition for the (101) anatase peak in grazing incident angle ($\omega = 5°$) is reported for 150 nm thick TaTO and $TiO_2$ films annealed with a standard vacuum (red and purple colored lines for TaTO and $TiO_2$ respectively) and an ultra-fast annealing (UFA) treatment in $N_2$ with 20 ppm of oxygen (blue and green colored lines for TaTO and $TiO_2$ respectively). *(b)* the total transmittance spectra for the same samples are reported (colors in agreement with (a)); the pink shaded part of the graph shows the visible region (400-700 nm) and the mean transmittance in the visible region ($T_{vis}$) is reported.

There is no significant difference in the crystallinity quality of TaTO and $TiO_2$ samples when annealed with the standard annealing or the UFA treatment, as indicated by the comparison between the most intense (101) XRD anatase signal (**Figure 3** *(a)*). The variation in both the intensity ratio and the absolute $2\theta$ positions of anatase X-ray peaks between TaTO and $TiO_2$ is consistent with a doping effect or with a defect-induced lattice distortion (or disordering), as discussed in reference[12]. A complete $\theta$-$2\theta$ scan (reported in the Supporting Information, **Figure s2**) does not show the presence of other $TiO_2$ polymorphs (e.g. rutile), or any other segregated phase (e.g. metallic Ta or $Ta_2O_5$). Moreover, the presence of $N_2$ in the annealing atmosphere (in both UFA and standard cycles) does not significantly hinder the crystallization of the thin films, since the anatase phase was formed in all cases.

Consistently with the XRD data, also the optical properties of doped and undoped anatase samples are shown to be independent of the employed thermal cycle. Quite remarkably, despite the significantly lower temperature and shorter time, the UFA treatment is not detrimental to the optical transparency of the crystallized thin films as shown by the superimposed transmittance curves for TaTO and $TiO_2$ after standard vacuum annealing (**Figure 3** *(b)*). In particular, the mean optical transmittance in the visible range of $TiO_2$ and TaTO reaches and exceeds the TCO technological limit of 80%.[29]

### 3.3 Oxygen Incorporation From the Crystallization Environment

In order to further study the effect of oxygen on the final electrical properties, we performed UFA treatments under nitrogen containing different $^{18}O$ concentrations. This allowed us to trace the oxygen penetration into TaTO thin films via depth profiles of the $^{18}O/^{16}O$ isotope ratio obtained with TOF-SIMS.

The *in-situ* electrical measurements recorded during UFA treatments performed in nitrogen atmosphere with 80 ppm and 1000 ppm of $^{18}O$ are reported in **Figure 4** *(a)* (purple and orange curves respectively). Note that the mixture of $N_2$ with 80 ppm of $^{18}O$ was used as it was the lowest $^{18}O$ concentration obtainable with our experimental setup. Let us consider region-A of Figure 4 *(a)* first: here it is possible to observe that the increase from 20 to 80 ppm of oxygen is already enough to affect the resistance behavior of the TaTO thin film. In line with the data discussed above (Figure 1), the further increase up to 1000 ppm of $^{18}O$ leads to an increased resistance, although the metallic behavior of the thin film is preserved. The resulting $R_{T=25°C}$ are 7 Ω and 10 Ω for 80 ppm and 1000 ppm $^{18}O$ respectively.

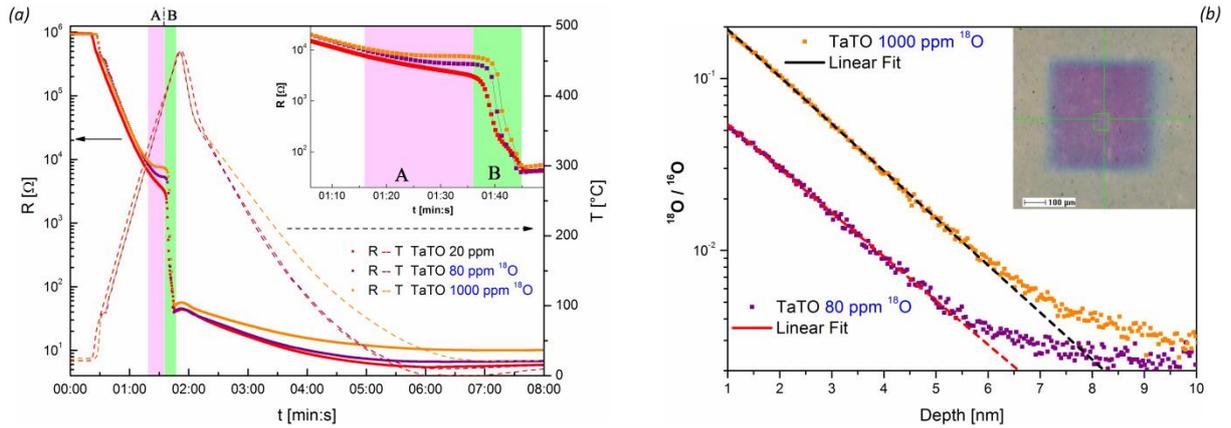

**Figure 4.** *(a)* In-situ resistance measurements (dots, left y-axes) and corresponding temperature cycles (dashed lines, right y-axes) for TaTO thin films crystallized in $N_2$-based atmospheres with different $^{18}O$ concentrations: 80 ppm (purple) and 1000 ppm (orange). TaTO thin film crystallized in $N_2$-based atmosphere with 20 ppm of oxygen (red) is reported as reference (already reported in Figure 1). The regions A and B (pink and green colored regions respectively) represent the time intervals in which the resistivity of the thin films starts to be affected by the presence of different oxygen concentrations (A) and in which the abrupt resistance drop takes place (B). In the inset a magnification of the resistance behavior in the regions A and B is reported. *(b)* the depth profile of the $^{18}O/^{16}O$ ratio traced via TOF-SIMS is reported; the colors of the dotted profiles are consistent with those used in *(a)*; in the inset is reported an optical microscope acquisition of the analyzed area of a TaTO thin film.

The TOF-SIMS results for TaTO samples annealed in $^{18}O$ are shown in **Figure 4** *(b)*. For both samples there is incorporation of oxygen. The decay of the $^{18}O/^{16}O$ ratio as a function of depth is well described by an exponential function in both cases (see the fitting lines in Figure 4 *(b)*). Note that while entering into the film, $^{18}O$ can also shift $^{16}O$ ions further into the sample, meaning that the $^{18}O$ profile does not fully correspond to the oxidation depth. Nonetheless, it is evident that UFA treatments performed under a higher concentration of $^{18}O$ (1000 vs. 80 ppm) resulted in (i) deeper $^{18}O$ penetration and (ii) larger $^{18}O$ concentration in proximity of the free surface of the TaTO film.

Although it is not possible to precisely determine the $^{18}O$ penetration depth one can still estimate that it should be limited to the first tens of nanometers of the thin films (Figure 4 *(b)*). Consistently with the $^{18}O$ SIMS profiles, this assumption was confirmed by sputtering removal of the topmost surface layers (tens of nm) from the UFA TaTO film crystallized in 1000 ppm (orange curve in Figure 1). Interestingly, the removal of the top 30 nm of the film resulted in the recovery of ohmic contact characteristics and led to an almost totally recovered resistivity value, which was characterized by the same charge carrier concentration with respect to TaTO film UFA in 20 ppm, ($\rho = 8.76 \times 10^{-4}$ $\Omega$cm, $n = 8.53 \times 10^{20}$ cm$^{-3}$). Note also that as the Ta concentration is so large (5 at.%) any possible increase of the oxygen vacancy concentration owing to the sputtering process should be negligible for the change of resistivity.

## 4. Discussion

A standard annealing treatment performed at ambient pressure in $N_2$ (grade 99.999%, oxygen concentration nominally < 3 ppm) results in insulating $TiO_2$-based thin films. This is the reason why in order to obtain highly conductive thin films a reducing atmosphere during a standard heat treatment is required. However, as shown above we found that during annealing a rather high content of oxygen can be tolerated as long as the heat treatment is extremely fast. Indeed, in terms of electrical properties, no significant differences (Table 1) were found upon ultra-fast-annealing treatments performed under reducing conditions (Ar/$H_2$ mixture, measured oxygen concentration < $10^{-20}$ ppm) or under rather oxidizing conditions (20 ppm $O_2$ in $N_2$) on TaTO (see Table 1).

Moreover, also the obtained structural (Figure 3 *(a)*), and optical properties (Figure 3 *(b)*) of all $TiO_2$-based thin films demonstrate the possibility of achieving practically the same thin film quality of a standard vacuum annealing process (with an overall process time of about 180 minutes) with a 5 minutes UFA treatment under nitrogen at ambient pressure. This finding has obviously important consequences not only for technological applications but also for a better understanding of the charge carrier chemistry of anatase.

From *in-situ* resistance measurements during UFA treatments, *ex-situ* analyses (e.g. Hall effect measurements) as well as SIMS data collected upon annealing in an $^{18}$O-containing atmosphere, it is evident that the oxygen exchange (incorporation) is closely related to the conductivity degradation of the $TiO_2$-based thin films during annealing.

In general, the oxygen exchange depends on temperature, oxygen partial pressure, and process time. Since here the annealing temperature was kept constant throughout all experiments, we varied $O_2$ content and annealing duration to investigate the role of oxygen incorporation. As summarized in Table 1, for the given $O_2$ concentration (20 ppm for the $N_2$ flux employed) and the peak temperature (460 °C) of the UFA treatment, we have observed no degradation of the electrical conductivity compared to (i) UFA treatments carried out under reducing conditions and, more importantly, (ii) to the standard vacuum annealing process. In contrast, extended UFA treatments (exposure to 460 °C for 10 min at 20 ppm $O_2$) led to the degradation of the electrical conductivity of both TaTO and $TiO_2$ thin films ($R_{T=25°C}$ = 11 Ω and 4965 Ω for TaTO and $TiO_2$ respectively) with respect to the conventional UFA cycle ($R_{T=25°C}$ = 6 Ω and 96 Ω for TaTO and $TiO_2$ respectively). It is important to emphasize that despite the rather low temperature (460 °C) and the limited duration of the heat treatment (10 min at 460 °C), the change of the electrical properties is considerable: a factor 2 increase of the resistance for TaTO and a factor 50 for $TiO_2$ (Figure s1). This further highlights the importance of the oxygen exchange between anatase thin films and the environment as far as the electrical properties are concerned.

Furthermore, it is worth noting that the very similar charge carrier density (see Table 1) obtained with a standard annealing cycle performed in vacuum and the UFA treatment at 20 ppm $O_2$ strongly indicates that upon PLD deposition the reducing conditions of the standard anneal do not create further oxygen vacancies (or at least not in a sufficient concentration to effectively change the concentration of mobile electrons), but rather that the oxygen-poor environment is useful for preventing oxygen incorporation during annealing. This is consistent with the possibility of tuning the charge carrier density of vacuum annealed TaTO thin films by adjusting the oxygen partial pressure during the room temperature PLD process as shown in our previous work (see Figure 6 in reference[12]).

As illustrated by XRD and optical microscopy data, during annealing the films become crystalline. In the light of the above considerations, we propose that under such conditions the amorphous $TiO_2$-based thin films undergoing the UFA treatment rapidly crystallize in the anatase phase without (or, in the case of oxygen concentrations > 20 ppm, with limited) oxygen incorporation from the surroundings, thus preventing the material from reaching equilibrium with the annealing atmosphere. This means in turn that the final thin film stoichiometry is in first approximation determined solely by the room temperature PLD deposition conditions. In this context, it is important to consider the presence of region-A in Figure 1, 2 and 4 *(a)*, (see Section 3.2), in which the resistance changes clearly depend on the oxygen partial pressure of the annealing environment for both doped and undoped $TiO_2$. We interpret the different variations of R with T in this region as an indication of oxygen incorporation occurring in the very first stages of the crystallization process. Previous studies performed on amorphous undoped and Nb-doped $TiO_2$ thin films already pointed out that the crystallization process can start with a sluggish rate at temperatures around 300 °C.[13, 27, 30-35] At higher temperatures ($T \geq 400$ °C), the crystallization process is faster, resulting in the abrupt drop of the resistance as indicated by the region-B in Figure 1 and Figure 2. The order of magnitude of the resistance drop in region-B for both $TiO_2$ and TaTO seems to be almost independent of the oxygen partial pressure, i.e. mainly related to the mobility increase, while going from amorphous $TiO_2$ to crystalline anatase (see Section 3.2). Consequently, this suggests that the oxygen incorporation during annealing is mostly effective in the first stage of the crystallization process (region-A), in which the crystallization rate of the process is still rather sluggish.

Let us consider now the UFA treatments at 1000 ppm $O_2$ and in air. In both cases, we observe that the detrimental effect on the electrical properties is notably different between doped and undoped samples. This can be rationalized in terms of defect chemistry: while moving from a reducing environment (PLD thin films deposited at room temperature) into a rather oxidizing one (UFA treatment performed at a rather high oxygen concentration) the increase of resistivity is expected to be more pronounced for the undoped composition, since the annihilation of oxygen vacancies ($V_O^{\bullet\bullet}$ in the Kröger-Vink notation) and eventually titanium interstitials ($Ti_i^{\bullet\bullet\bullet\bullet}$) could severely reduce the electron concentration of the undoped material.[36-39] In the case of Ta-doped TiO$_2$ instead, the high doping level (5 at.%) pins the electron concentration over a broad range of oxygen partial pressure ($n \sim 1.4 \times 10^{21}$ cm$^{-3}$ under the hypothesis of 100% Ta replacing Ti - $Ta_{Ti}^{\bullet}$) making the exposure to rather oxidizing conditions less problematic in terms of conductivity if an association/interplay among different defects is not taken in consideration. Nonetheless, in the donor-doped case, the concentration of negatively charged defects such as titanium vacancies ($V_{Ti}''''$) and oxygen interstitials ($O_i''$) is expected to be larger than in the undoped situation so that they might also contribute to decreasing the concentration of mobile electrons.[24, 25, 36, 37, 40]

SIMS data acquired upon $^{18}$O incorporation experiments prove that the oxygen insertion occurring during UFA at moderate O$_2$ concentration (80 ppm or 1000 ppm) is essentially limited to approximately the first tens of nm of the TaTO thin films (see Figure 4 *(b)*). This is nicely and independently confirmed by the removal of the surface of the thin films by sputtering: as the top 30 nm of the TaTO film were removed, the electrical conductivity was recovered and it was possible to verify that the charge carrier density matched the thin films subjected to UFA under 20 ppm O$_2$. We have thus experimentally demonstrated the presence of a thin surface layer on top of the highly conducting film, which is characterized by a higher oxygen concentration and resistivity. This eventually leads to a non-ohmic contact behavior in the case of TaTO thin films, which were UFA-crystallized at 1000 ppm and 21% oxygen concentration. It is noteworthy that the equivalent samples crystallized under the same UFA conditions with the evaporated Ti/Au electrodes on top exhibit an ohmic behavior between the electrode pads. This fact is consistent with oxygen penetration (and associated conductivity reduction) limited to the topmost uncovered surface of the thin film, while oxygen cannot reach the surface of the thin film underneath the Ti/Au evaporated contacts which therefore remains highly conducting.

The collected evidence is in line with the above discussed TaTO defect chemistry: the exposure of the material to rather oxidizing annealing conditions should (i) fill the existing oxygen vacancies (which are supposed to be present in the pristine thin films because of the reducing conditions employed during the room temperature PLD deposition) and (ii) possibly enhance the concentration of other defects such as titanium vacancies and oxygen interstitials, which can also act as 'electron killers'.[41] It is consequently reasonable to assume a significant electron density reduction in the thin surface layer induced by the environmental oxygen which is incorporated during the crystallization process, while the thin film beneath it preserves its high conductivity (and its stoichiometry). From the technological point of view, the presence of a thin layer characterized by a lower charge carrier concentration on top of the TCO layer is usually the basic condition required in a solid state dye sensitized/perovskite-based solar cell, in which the presence of a low charge carrier density selective layer is needed on top of the TCO in order to avoid high recombination rates between the photogenerated charge carriers.

Note that although the incorporation of nitrogen ions during the investigated UFA treatments is very unlikely (for this reason nitriding processes are performed in $NH_3$ or N-containing plasma rather than in $N_2$), nitrogen entering into anatase could substitute oxygen and act as an acceptor ($N'_O$). This would further reduce the concentration of electrons, in addition to the effect of the oxygen partial pressure which has been shown to be the ruling mechanism.[42]

Finally, it is noteworthy how fast the crystallization process proceeds leading to the formation of anatase grains with a typical lateral size of several micrometers within only few minutes. This is due to the characteristic 'explosive' crystallization of $TiO_2$.[12, 34, 35] This phenomenon has been attributed to the latent heat released during $TiO_2$ crystallization, which is large enough to result in a runaway process that continues until the amorphous material is completely consumed.[34] In our study, the lateral size distribution of the grains for both $TiO_2$ and TaTO thin films was found to be on the order of 10 μm for both the standard annealing and the UFA treatments (actually, for the standard treatment a larger average size is observed as shown by the optical microscopy images with polarized light reported in **Figure s3** in the Supporting Information).

## 5. Conclusion and Perspectives

We have demonstrated how ultra-fast annealing (UFA) treatments can be used to crystallize $TiO_2$-based thin films in the presence of mildly oxidizing conditions with an overall process time of just 5 minutes.

In particular, our experimental findings show that UFA treatments performed at ambient pressure in $N_2$ atmosphere (measured oxygen concentration of 20 ppm) allow for obtaining $TiO_2$-based TCOs with excellent electrical conductivity and transparency. Actually, the resulting properties are almost identical to those obtained through standard annealing treatments performed under a reducing atmosphere (e.g. vacuum). This is due to the possibility of avoiding and/or limiting (in a controlled way) oxygen incorporation during ultra-fast annealing.

Finally, this process is potentially highly appealing for the fabrication of new generation photovoltaic solar cells, in which $TiO_2$ already plays a key role (i.e. photoanode / selective layer). By engineering the UFA crystallization process as a function of the oxygen concentration in the annealing atmosphere and by finely controlling the oxygen penetration depth, it is possible to obtain TaTO films with tunable electrical properties as a function of the film depth. This reveals the possibility of fabricating an all-TaTO electrode, i.e. a TaTO TCO film with a top selective layer created by the UFA process, in which it is possible to tune the thickness of the selective layer as a function of the chosen device architecture in a single deposition followed by the annealing process.[18] This capability could reduce the number of sharp interfaces among different materials (e.g. FTO – TiO$_2$) in several solar cell configurations and consequently the possible presence of recombination centers[43] or energy barriers[11] for a more efficient electron collection.

**Supporting Information**
Supporting Information is available from the Wiley Online Library or from the author.

**Acknowledgements**
The authors wish to thank Rotraut Merkle for preparing the $^{18}$O/$^{16}$O mixtures and for helpful discussions, Udo Klock for the fundamental contribution in the realization of the ultra-fast annealing furnace, Andrea Ballabio for the current-voltage sweep measurements, Gennady Logvenov and Georg Christiani for the useful discussion on resistivity measurement of multi-layer structured thin films, and David Dellasega for the helpful discussion on the surface sputtering experiments.

Received: ((will be filled in by the editorial staff))
Revised: ((will be filled in by the editorial staff))
Published online: ((will be filled in by the editorial staff))

**We report on an ultra-fast crystallization process of transparent and conducting TiO$_2$ and Ta-doped TiO$_2$ thin films in nitrogen atmosphere at ambient pressure.** *In-situ* resistance measurements in controlled annealing atmospheres demonstrate the possibility to finely control their electrical properties in an extremely fast process highly appealing for new generation solar cell devices.

*P. Mazzolini, T. Acartürk, D. Chrastina, U. Starke, C.S. Casari, G. Gregori\* and A. Li Bassi\**

**Controlling the Electrical Properties of Undoped and Ta-doped TiO$_2$ Polycrystalline Films via Ultra-Fast Annealing Treatments Title**

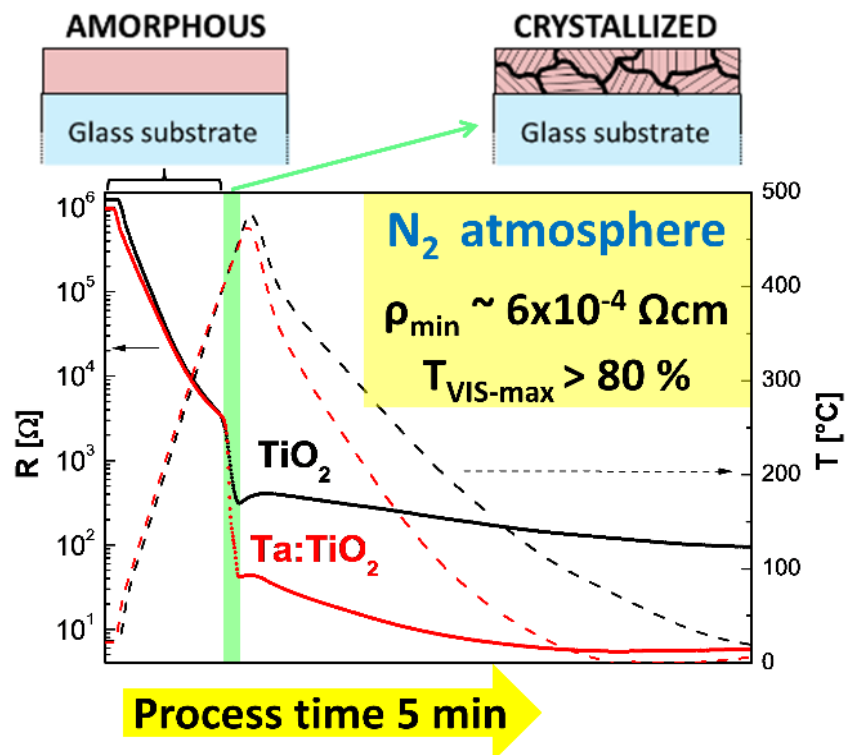

# Supporting Information

**Controlling the Electrical Properties of Undoped and Ta-doped TiO$_2$ Polycrystalline Films via Ultra-Fast Annealing Treatments**

*P. Mazzolini, T. Acartürk, D. Chrastina, U. Starke, C.S. Casari, G. Gregori\* and A. Li Bassi\**

**1. In-situ recorded temperature and resistance measurements during extended UFA (exposure to 460 °C for 10 min) treatments in N$_2$ (20 ppm O$_2$ concentration) for TiO$_2$ and TaTO thin films**

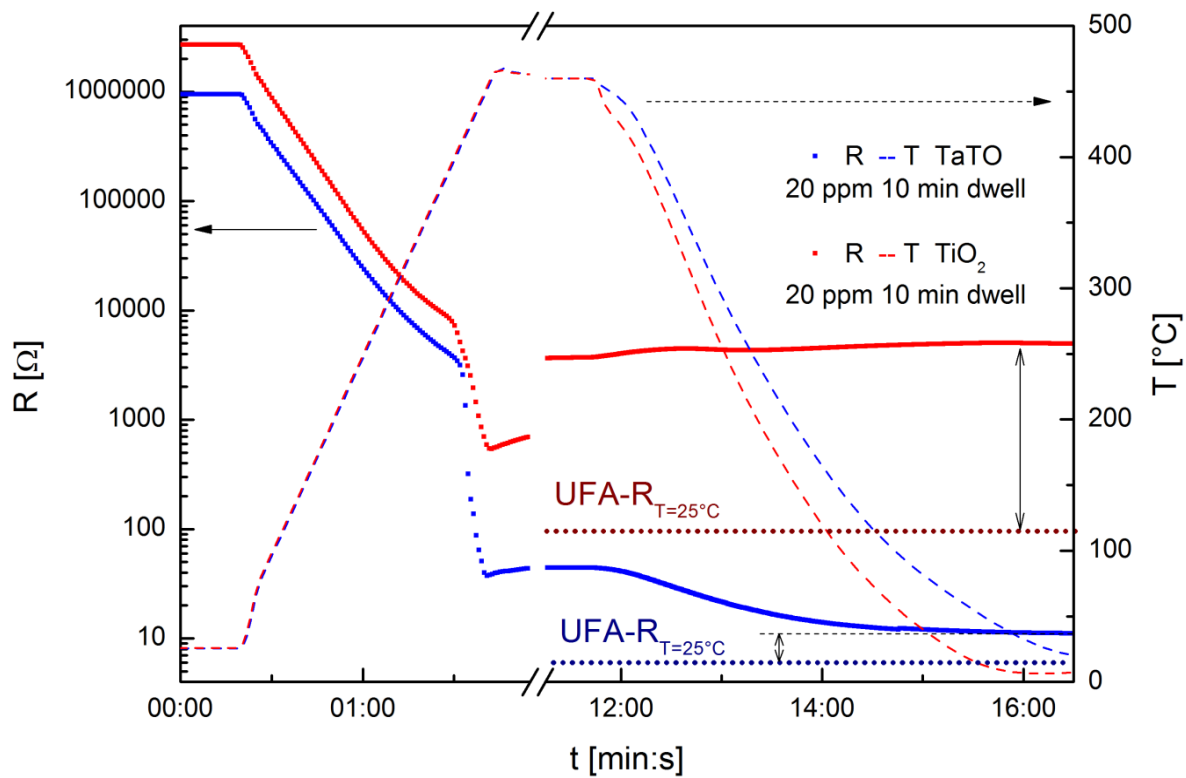

**Figure s1.** In-situ resistance measurements (dots, left y-axes) and corresponding temperature cycles (dashed lines, right y-axes) for TaTO (blue curves) and TiO$_2$ (red curves) thin films crystallized in N$_2$ (20 ppm O$_2$ concentration) with extended (10 minutes dwell at the peak $T$ = 460 °C) UFA cycles. The dotted horizontal lines reported after the axes brake (adopted in the dwell time laps for a clearer data presentation) are representing the resistance value obtained at room temperature ($R_{T=25°C}$) for TaTO (blue) and TiO$_2$ (red) thin films crystallized in UFA cycles performed in the same N$_2$ atmosphere without employing dwell time (complete acquired curves reported in Figure 1 and 2 for TaTO and TiO$_2$ respectively).

**2. X-Ray Diffraction (XRD) patterns of undoped and Ta-doped (TaTO) TiO$_2$ thin films crystallized with standard vacuum and N$_2$ ultra-fast annealing (UFA) cycle**

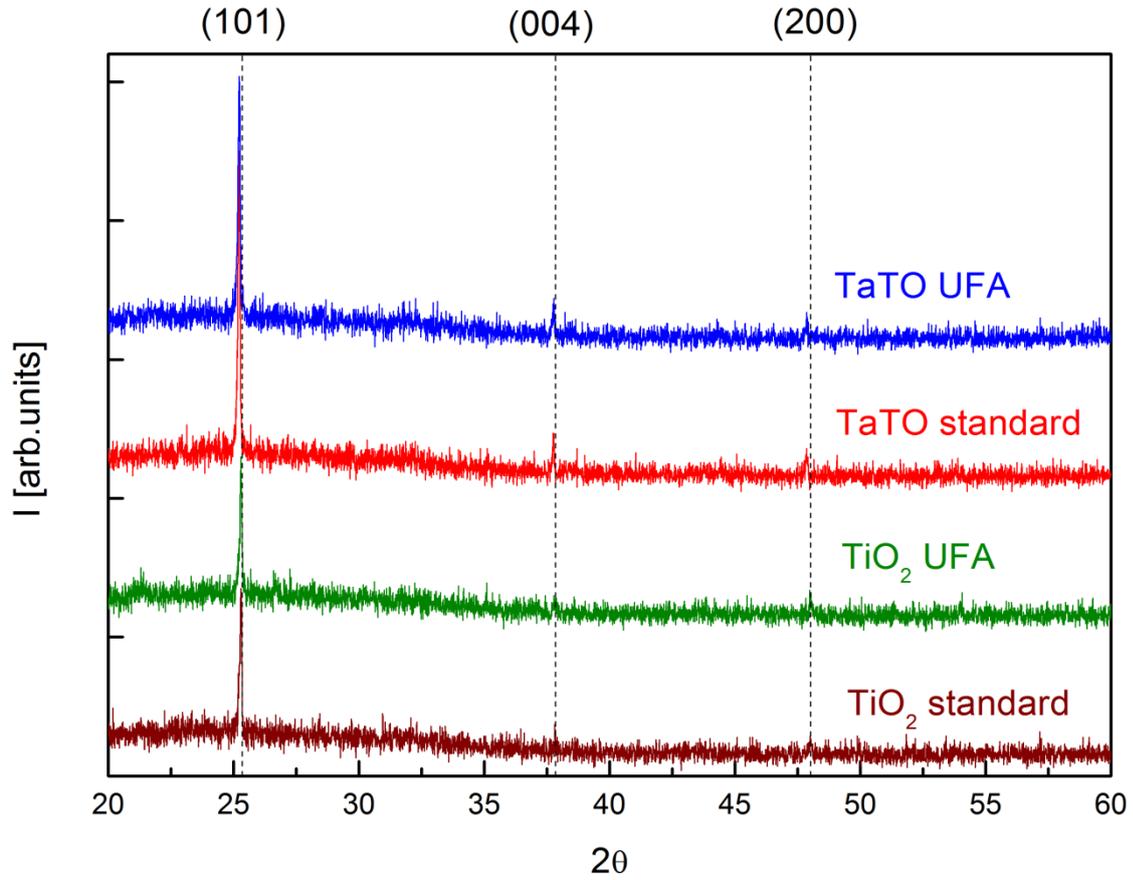

**Figure s2.** XRD acquisition in grazing incident angle ($\omega = 5°$) for 150 nm thick TaTO and TiO$_2$ films annealed with a standard vacuum (red and wine colored lines for TaTO and TiO$_2$ respectively) and an UFA treatment in N$_2$ (20 ppm oxygen concentration). In the upper part of the graph, the corresponding anatase diffraction plane indexes are reported.

**3. Optical microscopy images captured using polarized light of TaTO and TiO$_2$ thin film surfaces treated with standard vacuum and UFA treatment performed in N$_2$ (20 ppm O$_2$ concentration)**

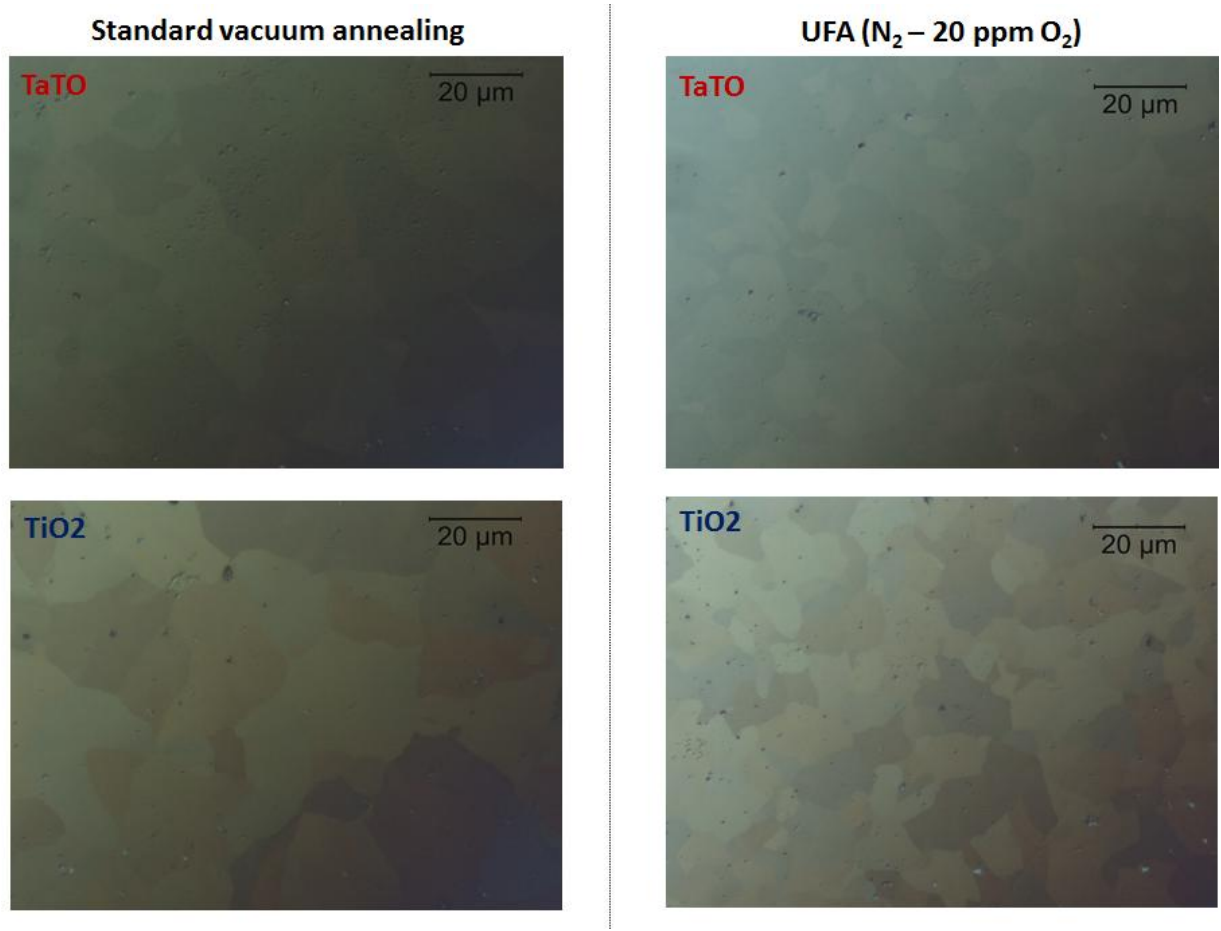

**Figure s3.** Surface images captured using polarized light through optical microscope of TaTO and TiO$_2$ thin films crystallized with a standard vacuum (left side of the figure) and UFA performed in N$_2$ atmosphere (right side of the figure).